\documentclass[aps,pre,twocolumn,float]{revtex4}
\usepackage{amsmath,bm,epsfig}
\usepackage{indentfirst}

\def\Fbox#1{\vskip1ex\hbox to 8.5cm{\hfil\fboxsep0.3cm\fbox{%
  \parbox{8.0cm}{#1}}\hfil}\vskip1ex\noindent}  

\newcommand{\B}[1]{{\bm{#1}}}
\newcommand{\C}[1]{{\mathcal{#1}}}    
\newcommand{\sFrac}[2]{{\textstyle\frac{#1}{#2}}}
\let \= \equiv \let\*\cdot \let\~\widetilde \let\-\overline

\begin{document}
\bibliographystyle{prsty}

\title{Predicting plastic flow events in athermal shear-strained amorphous solids}
\author{Smarajit Karmakar$^1$, Anael Lema\^{\i}tre$^2$, Edan Lerner$^1$ and Itamar Procaccia$^1$}
\affiliation{$^1$ Department of Chemical Physics, The Weizmann
Institute of Science, Rehovot 76100, Israel\\
$^2$ UMR Navier, 2 all\'e K\'epler, 77420 Champs-sur-Marne, France}
\begin{abstract}
We propose a method to predict the value of the external strain where a generic amorphous solid will fail by a plastic response (i.e. an irreversible deformation), solely on the basis of measurements of the nonlinear elastic moduli.  While usually considered fundamentally different, with the elastic properties
describing reversible phenomena and plastic failure epitomizing irreversible behavior, we show that the knowledge
of some nonlinear elastic moduli is enough to predict where plasticity sets in.
\end{abstract}
\maketitle

{\bf Introduction}: Studies of plasticity in amorphous solids
have always been hampered by the lack of any method
to identify {\it a-priori} the locations of dissipative (plastic)
events. This is in sharp contrast with crystals, where
plasticity can be assigned to the motion of identifiable,
topological defects like dislocations, the discovery of which
has triggered tremendous theoretical breakthroughs. In
the absence of any clear-cut definition of ``shear transformation
zones'', studies of amorphous systems are thus
still in search for reliable predictors of yielding.
Obvious observables such as local stress or density have
proven unreliable~\cite{81SMTEV,08TTLB}. More promising are studies of local
elastic fluctuations~\cite{04YJWNP,05YPLP,09TTGB,us}. It appeared that plastic
failure correlates with soft elastic regions. This could be expected
as yielding involves the crossing of
saddle points at which the shear modulus vanishes before
it presents a singular behavior~\cite{MaloneyLemaitre2004a}.
Nevertheless the values of the linear elastic
moduli alone do not carry predictive power; we
cannot just say that failure will occur in any given softer region in
space. The search for {\em reliable} predictors of plasticity
has thus become a major issue in studies of amorphous solids.
In this Letter we will show that an accurate predictor of plastic
failure in an athermal amorphous solid can be constructed,
as soon as higher order derivatives of the potential function
are involved. Our findings not only offer a predictive
tool for the onset of failure, but also point out the
importance of nonlinearities, and in particular those that couple
nonlinearly ``softening'' regions with strain at larger
scales.

\begin{figure}
 \centering
\includegraphics[scale = 0.31]{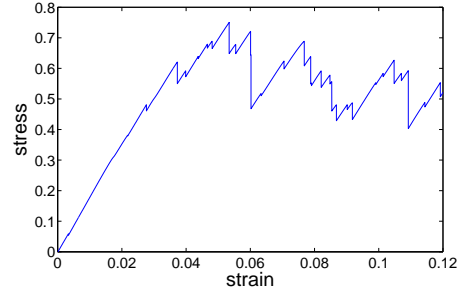}
\caption{A typical stress vs. strain curve in a system of 4096 particles in two dimensions obtained in the
athermal limit. Every elastic (reversible) increase in stress if followed by a sudden plastic (irreversible) drop
in stress. The aim of this Letter is to predict the value of the strain at which the next plastic drop will take place.}
\label{stress-strain}
\end{figure}

To fix ideas, imagine a simple shear deformation applied to a given piece of amorphous solid (for simplicity in 2D, with immediate extensions to 3D). A small strain increment $\delta\gamma$ corresponds to a change of the $i$'th particle positions $\B r_i\to \B r'_i$ as:
$
x'_i = x_i + \delta\gamma y_i $,
$y'_i  = y_i$ ,
In athermal quasi-static conditions ($T\to 0, \quad \dot\gamma\to 0$), the system lives in local minima, and follows strain-induced changes of the potential energy surface~\cite{98ML,MaloneyLemaitre2004}. Therefore, the particles do not follow homogeneously the macroscopic strain, and their positions change as $\B r_i\to \B r'_i+\B u_i$, where $\B u_i$ denotes non-affine displacements. Around some stable reference state at $\gamma=\gamma_0$, the field $\B u_i$, the system energy, and internal stress $\sigma_{xy}$ are smooth functions of $\gamma$. We can thus write:
\begin{equation}
\sigma_{xy} (\gamma) \!= \!\sum_{n=0}^\infty \frac{B_n}{n!} (\gamma-\gamma_0)^n\ , ~~ B_n =\lim_{T\to 0}\left. \frac{d ^n \sigma_{xy}}{d \gamma^n}
\right|_{\gamma = \gamma_0}. \label{series}
\end{equation}
As the strain increases, the system must eventually lose mechanical stability;  the ``elastic branch'' on the stress curve ends in a discontinuity as the system fails via a first subsequent ``plastic event'', see Fig. \ref{stress-strain}.  It is precisely at this instability, say at $\gamma=\gamma_P$, that the function $\sigma_{xy}(\gamma)$ loses its analyticity. Accordingly we recognize that the radius of convergence of the series (\ref{series}) is precisely $|\gamma_P-\gamma_0|$, where $\gamma_P$ can be larger or smaller than $\gamma_0$.

Our method to predict where plasticity sets in rests on two ideas. The first is that the coefficients in Eq. (\ref{series}) contain relevant information about the value of $\gamma$ where analyticity is lost. This comment is reminiscent of high temperature expansions in critical phenomena, where the knowledge of a substantial number of derivatives (and customarily using some Pad\'e resummation) can shed important light on the nature of the critical point \cite{74Dom}. The second is that, although we do not have access to a large number of derivatives (see below), we actually {\em know}~\cite{MaloneyLemaitre2004a} the nature of the singularity at $\gamma_P$, and we can use this knowledge to reach predictability which otherwise would be out of hand. Recall that as long as the system remains in mechanical equilibrium (i.e. along the
elastic branch) the force $\B f_i$ on every particle are zero before and after an infinitesimal deformation;
In other words~\cite{MaloneyLemaitre2004a,LemaitreMaloney2006} with $U$ the potential energy
\begin{equation}
\frac{d\B f_i}{d\gamma}=-\frac{d}{d\gamma}
\frac{\partial U}{\partial \B r_i}=-\frac{d}{d\gamma}
\frac{\partial U}{\partial \B u_i}=0,
\end{equation}
which implies
\begin{equation}
\frac{\partial^2 U}{\partial\gamma\partial \B u_i}+\frac{\partial^2 U}{\partial\B u_j\partial \B u_i}
\frac{d\B u_j}{d\gamma}
\equiv \B \Xi_i +\B H_{ij}\frac{d\B u_j}{d\gamma}=0 \ .
\end{equation}
This condition introduces the all-important Hessian matrix $\B H_{ij}$ and the `non-affine force' $\B \Xi_i$ which
can both be computed from the interparticle interactions. We rewrite this condition as
\begin{equation}
\frac{d\B u_i}{d\gamma}\!=\!-\B H_{ij}^{-1} \B \Xi_j \!=
-\!\!\sum_k\frac{ \B \psi^{(k)}_j\cdot \B \Xi_j }{\lambda_k}\B \psi^{(k)}_i\!\approx \! -\frac{ \B \psi^{(P)}_j\cdot \B\Xi_j }{\lambda_P}\B \psi^{(P)}_i \ , \label{duidt}
\end{equation}
where the second equation results from expanding in the eigenfunctions of $\B H$, $\B H_{ij} \B \psi_j^{(k)} =\lambda_k
\B \psi_i^{(k)}$; the last estimate stems from our knowledge that in finite systems the plastic event is associated with a single
eigenvalue going through zero when the systems slides over a saddle. Denote the critical eigenvalue as $\lambda_P$.
Eq. (\ref{duidt}) integrates to provide the distance of the non-affine field $\B u_i$ from its value at $\gamma_P$,
$
\B u_i(\gamma) - \B u_i(\gamma_P) = X(\gamma) \B \psi^{(P)}_i $,
where $X(\gamma)$ is a function of $\gamma$ only, satisfying $X(\gamma_P)=0$ and
\begin{equation}
\frac{dX(\gamma)}{d\gamma} \approx -\frac{ \B \psi^{(P)}_j\cdot \B\Xi_j }{\lambda_P}\B   \ . \label{dxdg}
\end{equation}
Finally, we use the crucial assumption~\cite{MaloneyLemaitre2004a} that the eigenvalue $\lambda_P$ crosses zero with a finite slope in the $X$-coordinate system itself, where distances are measured along the unstable direction:
\begin{equation}
\lambda_P \approx A X +{\C O}(X^2) \ , \label{lamX}
\end{equation}
Together with Eq.~(\ref{dxdg}) and asserting that $\B \Xi_j$ is not singular (it is a combination of derivatives of the potential function~\cite{LemaitreMaloney2006}), implies that
\begin{equation}
X(\gamma)\propto \sqrt{\gamma_P-\gamma} \ . \label{Xgam}
\end{equation}

These results are now used to determine the singularity of the stress at $\gamma_P$. We start with the exact result
for the shear modulus~\cite{MaloneyLemaitre2004a,LemaitreMaloney2006}
\begin{equation}
\mu =\mu_B -\B\Xi \cdot{\B H}^{-1}\cdot \B\Xi/V \ ,
\end{equation}
where $\mu_B$ is the Born term. Using Eqs.(\ref{lamX}) and (\ref{Xgam}) we conclude that near $\gamma_P$ we can write
the shear modulus as a sum of a regular and a singular term,
\begin{equation}
\mu \approx \tilde \mu - \frac{a/2}{ \sqrt{\gamma_P-\gamma}} + {\C O} ( \sqrt{\gamma_P-\gamma}) \ .
\end{equation}
 We thus assert that in the vicinity of $\gamma_P$:
\begin{equation}
\sigma_{xy}\sim \sigma_P+ a\sqrt{\gamma_P-\gamma}+b (\gamma_P-\gamma)^{3/2}+\cdots \ , \quad \gamma<\gamma_P \ .
\label{sigmap}
\end{equation}
Indeed, the stress begins to go down before the plastic even takes place, but this is not seen in Fig.~\ref{stress-strain} since this happens very sharply.
We can now come back to the question of predicting plasticity by looking at derivatives at any given point $\gamma_0<\gamma_P$. Near $\gamma_0$, we know that $\sigma_{xy}\sim \mu (\gamma-\gamma_0)+\dots$. Using~(\ref{sigmap}) we write the ansatz
\begin{equation}
\sigma_{xy} (\gamma) = \sigma_0 +\mu (\gamma-\gamma_0)+ a\sqrt{\gamma_P-\gamma}+b (\gamma_P-\gamma)^{3/2} \ , \label{ansats}
\end{equation}
from which we can recalculate the derivatives at $\gamma=\gamma_0$:
\begin{eqnarray}
B_2 &=&\frac{1}{4\sqrt{\gamma_P-\gamma_0}}\left[3b -\frac{a}{\gamma_P-\gamma_0}\right]\ , \label{Bn}\\
B_3&=&\frac{3}{8(\gamma_P-\gamma_0)^{3/2}}\left[b -\frac{a}{\gamma_P-\gamma_0}\right]\ , \nonumber\\
B_4&=&\frac{3}{16(\gamma_P-\gamma_0)^{5/2}}\left[3b -\frac{5a}{\gamma_P-\gamma_0}\right]\ .
\end{eqnarray}
Note that the shear modulus has disappeared from these expressions, becoming irrelevant for the subsequent prediction of the instability threshold. By measuring these three derivatives at $\gamma_0$ we can determine all the unknowns in Eq.(\ref{ansats}). $\gamma_P$ solves a 4th order polynomial, and
\begin{equation}
\gamma_P = \gamma_0+ \frac{3B_3-\sqrt{9B_3^2-2B_2B_4}}{2B_4} \ , \label{solution}
\end{equation}
is the only physical solution among the four available ones. Next we test these results in a specific model.

{\bf Model and numerical procedures}:
Below we employ a model system with point particles of two `sizes' but of equal mass $m$
in two-dimensions, interacting via a pairwise potential of the form
\begin{equation}\label{potential}
\phi\left(\!\frac{r_{ij}}{\lambda_{ij}}\!\right) =
\left\{ \begin{array}{ccl} \!\!\varepsilon\left[\left(\frac{\lambda_{ij}}{r_{ij}}\right)^{k} + \displaystyle{\sum_{\ell=0}^{q}}c_{2\ell} \left(\frac{r_{ij}}{\lambda_{ij}}\right)^{2\ell}\right] &\! , \! & \frac{r_{ij}}{\lambda_{ij}} \le x_c \\ 0 &\! , \! & \frac{r_{ij}}{\lambda_{ij}} > x_c \end{array} \right., \end{equation} where $r_{ij}$ is the distance between particle $i$ and $j$, $\varepsilon$ is the energy scale, and $x_c$ is the dimensionless length for which the potential will vanish continuously up to $q$ derivatives. The interaction lengthscale $\lambda_{ij}$ between any two particles $i$ and $j$ is $\lambda_{ij} = 1.0\lambda$, $\lambda_{ij} = 1.18\lambda$ and $\lambda_{ij} = 1.4\lambda$ for two `small' particles, one `large' and one `small' particle and two `large' particle respectively. The coefficients $c_{2\ell}$ are given by \begin{equation} c_{2\ell} = \frac{(-1)^{\ell+1}}{(2q-2\ell)!!(2\ell)!!}\frac{(k+2q)!!}{(k-2)!!(k+2\ell)}x_c^{-(k+2\ell)}.
\end{equation}
We chose the parameters
$x_c = 7/4$, $k=10$ and $q=6$.
The unit of length $\lambda$ is set to be the interaction length scale of two small particles, and
$\varepsilon$ is the unit of energy.
The density for all systems is set to be $N/V = 0.85\lambda^{-2}$. We employ an athermal quasi-static scheme which consists of imposing an affine transformation to each particle of a configuration, followed by a potential energy minimization under Lees-Edwards boundary conditions \cite{91AT}. In this scheme one can obtain purely elastic trajectories of stress vs strain \cite{09LP}, which allows for the calculation of the total derivatives of stress with respect to strain using finite differences; we choose the stopping criterion for the minimizations to be $|\nabla_iU| < 10^{-25}\frac{\varepsilon}{\lambda}$ for every coordinate $x_i$, and select the strain increment for taking derivatives to be $\delta\gamma =5\times 10^{-7}$.

It is useful to compare our prediction with a simple failure criterion which uses a naive Taylor series of the form
$\sigma_{xy}(\gamma) = \sigma_0 + \mu(\gamma-\gamma_0) + \frac{1}{2}B_2(\gamma-\gamma_0)^2
+\frac{1}{6}B_3(\gamma-\gamma_0)^3+\frac{1}{24}B_4(\gamma-\gamma_0)^4$.
From here $\gamma_P$ is estimated as the point at which
$|\frac{d\sigma}{d\gamma}|_{\gamma=\gamma_P} = 0$--the difference between this point and the actual yield point is negligible compared to all other strain scales entering the problem \cite{foot}. This is the solution of the equation:
\begin{equation}
\mu + B_2(\gamma_P-\gamma_0) + \sFrac{1}{2}B_3(\gamma_P-\gamma_0)^2 + \sFrac{1}{6}B_4(\gamma_P-\gamma_0)^3 = 0\ .  \label{Taylor2}
\end{equation}
 \begin{figure}
 \centering
\includegraphics[scale = 0.25]{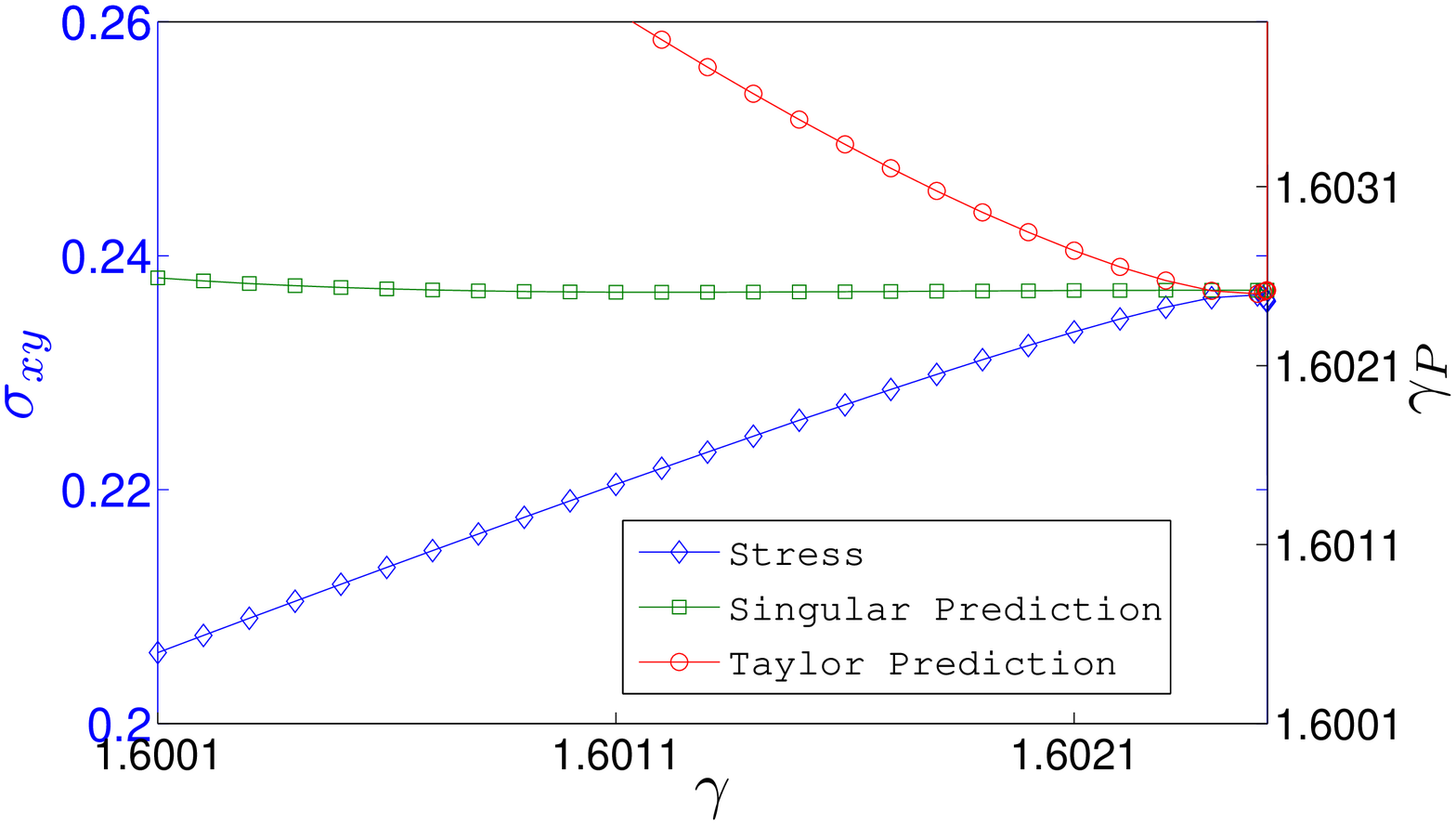}
\includegraphics[scale = 0.25]{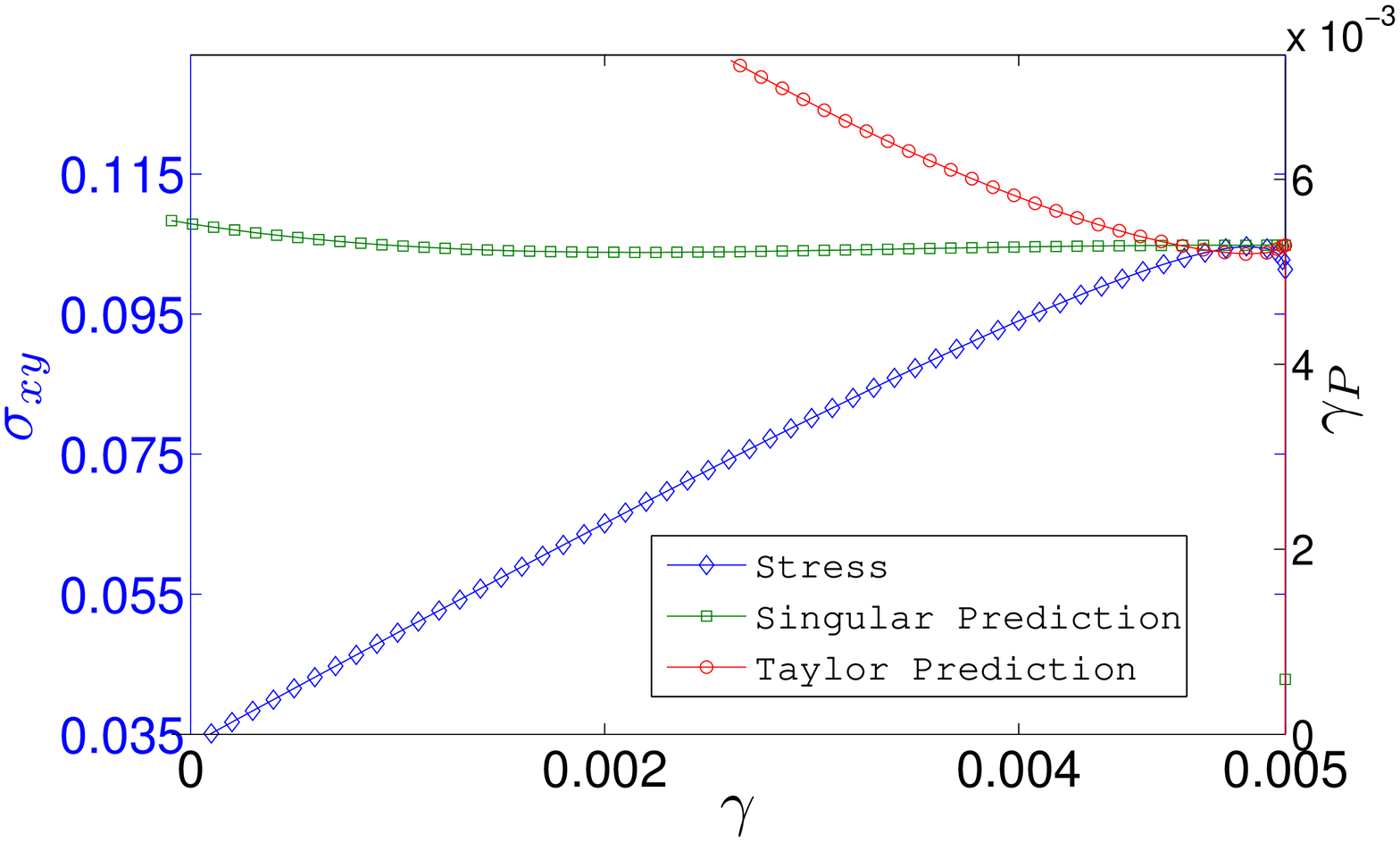}
\caption{Color online. Upper panel: Straining from an elastic branch of the elasto-plastic steady state in a system of $N=484$ particles.
Shown are the stress vs. strain curve up to
the first plastic event (blue rhombi) and the value of $\gamma_P$ as predicted by Eq. (\ref{solution}) for every value of $\gamma$ (green squares).
In addition we show the prediction of the simplistic Taylor expansion (\ref{Taylor2}) (red circles). Lower panel: the same measurements starting from the equilibrium isotropic stress-less state. Note the region where the
even derivatives build up.}
 \label{prediction}
 \end{figure}

{\bf Predictions}: Fig. \ref{prediction} in the upper panel demonstrates the quality of our predictor for a typical initial condition taken in the elasto-plastic steady state in a system with $N=484$.
Here, $\gamma_0$ corresponds to the strain at which the elastic branch was reached via a prior plastic event. The total length of the branch on the stress-strain curve is thus exactly $\gamma_P-\gamma_0$. We show the stress vs. strain curve in (symbols) and the values of $\gamma_P$ predicted using either Eq. (\ref{solution}) or (\ref{Taylor2}) at each value of the external strain.
We see that the naive Taylor expansion fails except right at $\gamma_P$, whereas our accelerated method works very well over nearly the whole length of the elastic branch. Of course, in the elasto-plastic steady state the strain range between plastic events is rather limited--as reflected by the scale of abscissa. But our data shows that the transition between one plastic event and the next in entirely controlled by a single mode, i.e. by a single soft zone. We will shortly come back on this issue.

We also test our predictor starting from an equilibrium isotropic state obtained by thermal annealing. Here we need to predict the {\em first} plastic event which occurs upon increasing the external strain. This situation could be expected to be more tricky since at $\gamma=\gamma_0=0$ all the even derivatives $d^{2n}\sigma_{xy}/d\gamma^{2n}$ vanish on the average. We could expect that, upon straining, these
derivatives must build up before predictability of plasticity is achieved. In fact, as illustrated on
 the lower panel of Fig. \ref{prediction} predictability is often achieved rapidly, even though the strain range is considerably larger than in the upper panel.
\begin{figure}
\centering
\includegraphics[scale = 0.37]{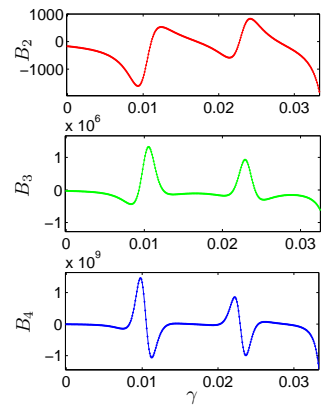}
\caption{The first three nonlinear derivatives $B_n$ as a function of $\gamma$, displaying the huge
variation in value due to large nonaffine reversible transformation before diving to $-\infty$ because
of a plastic event.}
\label{nonaffine}
\end{figure}
In steady state the vast majority of the elastic branches end up in a plastic event whose
onset can be predicted as demonstrated here. In contradistinction, the elastic branch emanating from the
isotropic equilibrium state may exhibit large non-affine elastic events which resemble a typical precursor to a plastic failure but avoids it by eventually stabilizing. Our derivatives will pick up these elastic events and will incorrectly predict a plastic failure before landing on the right prediction. This scenario is shown in Fig. \ref{nonaffine} in which two non-affine large events occur before the first plastic event. All three derivatives undergo huge changes at
the non-affine events and then dive to $-\infty$ at the true plastic event. This seems to be a trace of the typical presence of several interacting eigenmodes, not necessarily well-aligned with the direction of shearing. We stress
that this phenomenon is quite rare and the normal situation is the one that is exhibited in Fig. \ref{prediction}
lower panel.

\begin{figure}
\centering
\includegraphics[scale = 0.435]{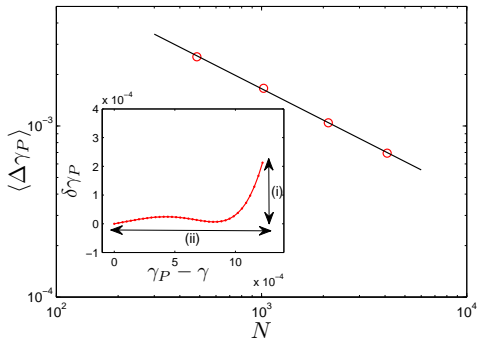}
\caption{ Log-log plot of the range of strain $|\gamma-\gamma_P|$ where $\delta \gamma_P/\Delta\gamma_P\le 0.15$ as a function of the system size in the elasto-plastic steady state. The slope of the continuous fit line is $\eta=-0.61$. The inset shows how the measurement
is done, $(i)/(ii)\le 0.15$. }
\label{accuracy}
\end{figure}

{\bf Dependence on system size}: The data shown in Figs. \ref{prediction} and \ref{nonaffine} pertain to a relatively small system of
484 particles. We therefore must raise the important question of how the predictability of the plastic failure depends on the system size. To study the size dependence of the predictability we
measured the distance $\Delta\gamma_P\equiv |\gamma-\gamma_P|$ for which the error in estimating $\gamma_P$, denoted as $\delta \gamma_P$, satisfies $\delta \gamma_P/\Delta\gamma_P\le 0.15$, for systems of varying sizes (see inset in Fig. \ref{accuracy}). The results are shown in Fig. \ref{accuracy}, indicating
that the range of predictability reduces like a power law with the system size,
\begin{equation}
\Delta\gamma_P \sim N^\eta\ , \quad \eta\approx -0.61 \ . \label{defeta}
\end{equation}
To estimate the exponent $\eta$ theoretically, we note that for predictability to be possible we need the divergent term of at least $B_2$ to be of the order of unity. The quantity $B_2$ (cf. Eq. (\ref{Bn})) has a coefficient $a$ which
is of order of $1/N$ because there is only a single localized mode that becomes unstable \cite{MaloneyLemaitre2004}, and the divergent term
that goes like $(\gamma-\gamma_P)^{-3/2}$. Thus we expect predictability when $|\gamma-\gamma_P|\sim N^{-2/3}$, which
estimates $\eta=-2/3$. Higher order derivatives are more singular and thus $\eta$ can only be smaller due to their effect. This estimate is very important, since it guarantees that predictability will not deteriorate
in the elasto-plastic steady-state. To see this, we recall that the average distance between successive plastic
events $\langle \Delta \gamma \rangle$ is known \cite{09LP} to follow a scaling law with the system size
$
\langle \Delta \gamma \rangle \sim N^\beta \ .
$
The exponent $\beta$ was measured in a variety of systems in 2D, and was found to be always in the range [-0.67, -0.63]. In the present model system we measure $\beta\approx -0.65$. With the present accuracy we cannot exclude that $\eta\approx \beta$, leading to the realization that while
the predictability is reduced with increasing $N$, the range of $\gamma$ over which we need to predict reduces
almost at the same speed, if not slightly faster. Thus in effect the predictability in the elasto-plastic steady state does not deteriorate with $N$.

In summary, we showed that (i) the discussion of plasticity in amorphous solids calls for understanding the role
of nonlinear elasticity, and (ii) for the elasto-plastic steady state the knowledge of a few nonlinear elastic constants suffices to predict where the next plastic event should occur. Finally, the analysis presented above is global, but it can be extended to local instability maps eliminating the system size dependence whatsoever. The first step toward a local analysis, which is the microscopic definition of the nonlinear elastic constants, has been already achieved and is available in Ref. \cite{10KLP}. The second step where it will be shown how to use these results in models of elasto-plasticity in both two and three dimensions will be discussed elsewhere \cite{10KLPa}.

\acknowledgments
This work had been supported in part by the Israel Science Foundation and the Ministry of Science under the French-Israeli collaboration.


\begin{thebibliography}{99}

\bibitem{81SMTEV}
D. Srolovitz, Maeda,Takeuchi, Egami and Vitek {\it et~al.}, J Phys-F-Metal Phys {\bf 11},  2209  (1981).

\bibitem{08TTLB}
M. Tsamados, A. Tanguy, F. Leonforte, and J.~L. Barrat, European Physical
  Journal E {\bf 26},  283  (2008).

\bibitem{04YJWNP}
K. Yoshimoto,T.~S. Jain, K.~V. Workum, P.~F. Nealey and J.~Pablo, Phys. Rev. Lett. {\bf 93},  175501  (2004).

\bibitem{05YPLP}
K. Yoshimoto, G.~J. Papakonstantopoulos, J.~F. Lutsko, and J.~J. de~Pablo,
  Physical Review B {\bf 71},  184108  (2005).

\bibitem{09TTGB}
M. Tsamados, A. Tanguy, C. Goldenberg, and J.~L. Barrat, Physical Review E {\bf
  80},  026112  (2009).

  \bibitem{us}
V. Ilyin, I. Procaccia, I. Regev, and N. Schupper, Phys. Rev E {\bf 77}, 061509 (2008).

  \bibitem{MaloneyLemaitre2004a}
C. Maloney and A. Lema\^{\i}tre, Phys. Rev. Lett. {\bf 93},  195501  (2004).


\bibitem{98ML}
D.~L. Malandro and D.~J. Lacks, Phys. Rev. Lett. {\bf 81},  5576  (1998).

\bibitem{MaloneyLemaitre2004}
C. Maloney and A. Lema\^{\i}tre, Phys. Rev. Lett. {\bf 93},  16001  (2004).

\bibitem{74Dom}
See for example: C. Domb in {\em Phase Transitions and Critical Phenomena} vol 3 ed C Domb and M S
Green (London: Academic Press) pp. 357–458 (1974).

\bibitem{LemaitreMaloney2006}
A. Lema\^{\i}tre and C. Maloney, J. Stat. Phys. {\bf 123},  415  (2006).

\bibitem{09LP} E. Lerner and I. Procaccia, Phys. Rev. E {\bf 79}, 066109 (2009).

\bibitem{91AT} M.P. Allen and D.J. Tildesley, {\em Computer Simultions of Liquids} (Oxford University Press, 1991).

\bibitem{foot}
The difference is $O(10^{-4})$ for $N=484$, and it decays like $N^{-2}$ for
larger systems, following the same arguments leading to the estiamte of $\eta$ in Eq.(\ref{defeta}).

\bibitem{10KLP}
S. Karmakar, E. Lerner, and I. Procaccia, ``Athermal Nonlinear Elastic Constants of Amorphous Solids",
Phys. Rev. E, submitted. Also: arXiv:1004.2198 .

\bibitem{10KLPa}
S. Karmakar, E. Lerner, and I. Procaccia, ``Local Nonlinear Elastic Constants: The Mechanism of Spatial
Correlations in Elasto-Plasticity of Amorphous Solids", in preparation.

\end{thebibliography}
\end{document}